\documentclass[aps,prl,tightenlines,twocolumn,showpacs,preprintnumbers,amsmath,amssymb]{revtex4}
\usepackage{graphicx}
\usepackage{dcolumn}
\usepackage{bm}
\usepackage{epsfig}

\begin{document}

\title{Probing the intrinsic Josephson potential in Bi-2212 by thermal activation}

\author{ V.M.Krasnov$^{1,2}$, T.Bauch$^2$, and P.Delsing$^2$}

\address{$^1$ Department of Physics, Stockholm University, Albanova University Center, SE-10691 Stockholm, Sweden\\
$^2$ Department of Microtechnology and Nanoscience, Chalmers
University of Technology, SE-41296 G\"oteborg, Sweden}


\begin{abstract}
We study thermal fluctuation phenomena in small Bi-2212 intrinsic
Josephson junctions. Being able to measure switching currents of a
{\it single} intrinsic junction, we observe that it's statistics
can be very well described by thermal activation from a periodic
Josephson potential with the sinusoidal current-phase relation.
This is a direct evidence for the dc-intrinsic Josephson effect
and the first unambiguous confirmation of the tunnelling nature of
interlayer transport in strongly anisotropic high temperature
superconductors. Furthermore, the fluctuation-free critical
current, extracted from the analysis of switching current
statistics, exhibits a temperature dependence typical for
superconductor- insulator- superconductor tunnel junctions.

\pacs{74.72.Hs
74.40.+k,
74.81.Fa
}
\end{abstract}
\maketitle

The nature of interlayer transport in high $T_c$ superconductors
(HTSC) has been a long standing question \cite{Interlayer}. It is
established that in extreme anisotropic Bi- and Tl-based HTSC
intrinsic Josephson effect exists between superconducting CuO$_2$
planes \cite{Kleiner}. Both dc and ac intrinsic Josephson effects
were observed in Bi$_2$Sr$_2$CaCu$_2$O$_{8+x}$ (Bi-2212), in the
form of flux quantization \cite{Fiske,LatyshPhC,Ooi}, Fiske
\cite{Fiske} and Shapiro \cite{Wang} steps in Current-Voltage
characteristics (IVC's) and the Josephson plasma resonance
\cite{Plasma}. On the other hand, the role of blocking Bi-layers
and the nature of interlayer coupling in Bi-2212 is still unclear,
partly due to difficulties in analysing intrinsic Josephson data
caused by stacking and strong electromagnetic coupling of
intrinsic Josephson junctions (IJJ's) \cite{Compar}.

The type of Josephson coupling can be deduced from the dependence
of the Josephson energy $E_J$ on the phase difference $\varphi$,
or, similarly, from the Josephson current-phase relationship
$I_s(\varphi)=(2e/\hbar)\partial E_J/\partial \varphi$, which
varies from a saw-tooth like for metallic weak links to the
sinusoidal for superconductor- insulator- superconductor (SIS)
tunnel junctions \cite{RevMP}. The $E_J(\varphi)$ determines the
junction electrodynamics, which is equivalent to motion of a
particle in a "tilted wash-board" potential, created by
superposition of the periodic Josephson potential $E_J(\varphi)$
and the work done by the current source, $-(\hbar/2e)I\varphi$. At
a finite temperature, $T$, the particle can escape from the
potential well as a result of thermal fluctuations, see Fig. 3 a).
This corresponds to switching of the junction from the
superconducting to the resistive state. The rate of thermal escape
\cite{Martinis,Grabert},
\begin{equation}
\Gamma_t(I) = a_t\frac{\omega_a}{2\pi} \exp\left[-\frac{\Delta
U}{k_B T}\right], \label{Rate}
\end{equation}
is a sensitive probe of the $E_J(\varphi)$ both via the attempt
frequency, $\omega_a$, {\it i.e.}, the frequency of oscillations
at the bottom of the potential well, and, particularly, via strong
exponential dependence of $\Gamma_t$ on the potential barrier
$\Delta U$. Thus, the switching current statistics carries direct
information about the shape of $E_J(\varphi)$ and, therefore,
about the nature of Josephson coupling.

In this letter we study the effect of thermal fluctuations in
small Bi-2212 intrinsic Josephson junctions. Being able to measure
switching currents of a {\it single} IJJ in an electrically
shielded environment, we observe that it's statistics can be very
well, and without fitting parameters, described by thermal
activation from the tilted wash-board potential with the
sinusoidal current-phase relation. This is a direct evidence for
the dc-intrinsic Josephson effect and the first unambiguous
confirmation of the tunnelling nature of interlayer transport in
Bi-2212. We also demonstrate that thermal fluctuations
dramatically affect properties of small IJJ's, resulting in strong
suppression of the switching current and unusual temperature
dependence in the whole $T-$range. However, the fluctuation-free
critical current $I_{c0}$, extracted from the analysis of
switching current histograms, exhibit $T-$dependence typical for
SIS junctions, consistent with tunnelling nature of the interlayer
coupling.

The IJJ's were fabricated by etching small mesa structures on top
of Bi-2212 single crystals. We developed a simple procedure
capable of fabricating deep sub-micron multi-terminal IJJ's. Fig.
1 shows a sketch of fabrication procedure. It involved
self-alignment cross-bar photolithography, during which an
insulating CaF$_2$ layer was formed using lift-off, see Fig. 1 a),
and $\sim 3 \times 3 - 5\times 5 \mu m^2$ mesas were formed in a
self-aligned manner, Fig. 1 c), at the crossing between the
narrow-long mesa, Fig. 1 a) and bar-like electrodes, Fig. 1 b).
Finally, the sample was transferred into standard Focused Ion Beam
(FIB) system (FEI Inc. FIB-200), and a smaller mesa was trimmed by
cutting of a portion of the mesa, as shown in Fig. 1 d). Due to
self-alignment at the previous stage there is no parasitic area
below the electrode and deep-submicron mesas can be fabricated.
Fig.1 e) shows IVC's of mesas before and after FIB trimming to
sub-micron dimensions. Both IVC's exhibit a knee at the sum-gap
voltage, followed by almost temperature independent high bias
resistance \cite{Krasnov_TH}, which is typical for SIS tunnel
junctions. The increase in resistance after trimming is in
agreement with $\sim 80$ fold decrease in the mesa area.

\begin{figure}
\noindent
\begin{minipage}{0.48\textwidth}
\epsfxsize=.9 \hsize \centerline{ \epsfbox{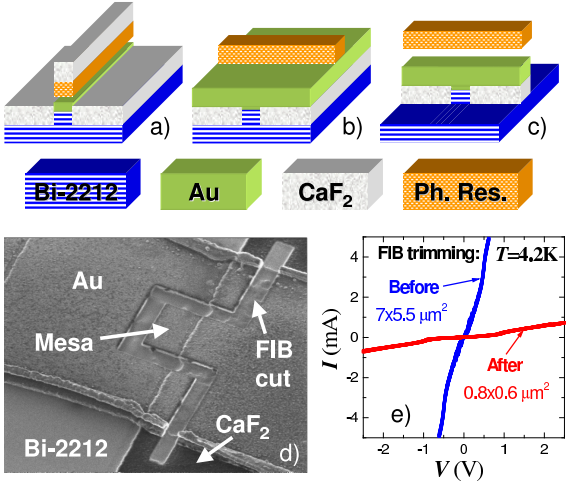} }
\caption{A sketch of sample fabrication: a-c) Self - alignment
cross-bar lithography, d) secondary electron image of the sample
after FIB trimming and e) I-V curves of mesas before and after FIB
trimming.}
\end{minipage}
\end{figure}

Fig. 2 shows IVC's, normalized by the mesa area, at $T \simeq 6 K$
for the initial $6 \times 3 \mu m^2$ mesa and two smaller mesas
obtained by cutting the initial mesa in two parts with the FIB.
The multi-branch structure of the IVC's is due to one-by-one
switching of the IJJ's from the superconducting to the resistive
(quasiparticle) state. Naively, it could be expected that the
normalized IVC's should collapse in one, since the critical
current density is a material property, independent of the
junction area. In reality, the measured switching current density
decreases with the mesa area. To understand whether this is caused
by deterioration of the IJJ's during FIB trimming and thermal
cycling, we checked the scaling of quasi-particle branches in the
IVC's. The thin solid lines in Fig. 2 represent multiple-integers
of the fit to the first branch of the initial mesa. It is seen
that quasiparticle branches of the initial mesa are perfectly
periodic. However, it is also seen that the normalized
quasiparticle branches of the trimmed mesas fall onto the same
fitting curves. This implies that there is no visible
deterioration of IJJ's during FIB trimming and thermal cycling.
Therefore, the observed decrease of the switching current density
in small mesas should be attributed to enhanced thermal
activations due to decrease of the potential barrier $\Delta U$,
which is proportional to the Josephson energy $E_{J0}=(\hbar
c/2e)I_{c0}$ and scales with the junction area.

\begin{figure}
\noindent
\begin{minipage}{0.48\textwidth}
\epsfxsize=0.9\hsize \centerline{ \epsfbox{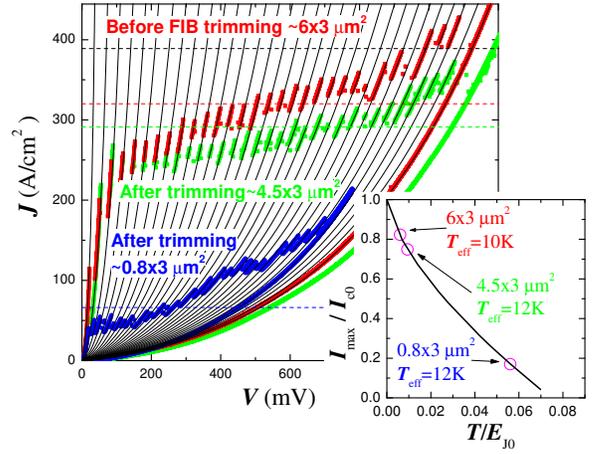} }
\caption{ Current density vs. Voltage curves for a mesa before and
after FIB cutting in two non-equal parts. The measured switching
current density decreases with the size of the mesa. Solid lines
are fits to quasiparticle branches, which are multiple integers of
the first branch. It is seen that quasiparticle branches of all
junctions fall onto the same line irrespective of the size of the
mesa, which implies that there is no junction deterioration.
Horizontal dashed lines show values of the fluctuation free
critical current density (the top line) and the most probable
switching current densities for the three mesas, obtained from the
fit shown in the inset. Inset: The solid line shows simulated
dependence of the most probable switching current $I_{max}$ vs.
$T/E_{J0}$, calculated for a classical thermal escape from a
tilted wash board potential. Circles indicate the best fit for the
there mesas. }
\end{minipage}
\end{figure}

For a quantitative analysis of the switching currents in small
mesas we consider the effect of thermal activation from a tilted
wash board potential. The probability to measure the switching
current $I$ is given by
\begin{equation}\label{P_I}
P(I)=\frac{\Gamma(I)}{dI/dt} \left[1-\int^I_0 P(I)dI\right],
\end{equation}
where $dI/dt$ is the current sweeping rate. For underdamped
junctions, $Q=\omega_pRC\gg1$, the prefactor in Eq.(\ref{Rate}) is
$a_t=4/\left[(1+Q k_B T/1.8 \Delta U)^{0.5} + 1\right]^2$
\cite{Grabert}. In simulations we will assume the sinusoidal
$I_s(\varphi)$, for which \cite{Martinis}
$\omega_a=\omega_p(1-(I/I_{c0})^2)^{1/4}$, where $\omega_p=(2e
I_{c0}/\hbar c C)^{1/2}$ is the Josephson plasma frequency, and
\begin{equation}
\Delta U \simeq \frac{4 \sqrt{2}}{3} E_{J0}\left[1-\frac I
I_{c0}\right]^{3/2}. \label{DU}
\end{equation}

\begin{figure}
\noindent
\begin{minipage}{0.48\textwidth}
\epsfxsize=.9\hsize \centerline{ \epsfbox{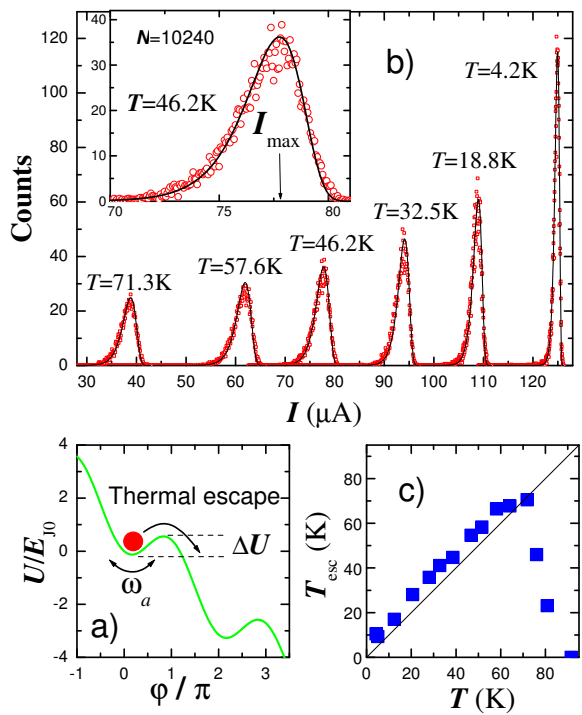} }
\caption{a) The shape of tilted wash-board potential for
$I/I_{c0}=0.5$. b) Switching current histograms of a single IJJ at
different temperatures. Solid lines represent fits to classical
thermal escape from the tilted wash-board potential. The inset
shows the quality of the fit for $T=46.2K$. Dots and lines
represent experimental data and fits, respectively. c) The
effective "escape temperature" vs. temperature. }
\end{minipage}
\end{figure}

The inset in Fig. 2 shows the simulated dependence of the most
probable switching current $I_{max}$ as a function of the ratio
$T/E_{J0}$, made for the typical experimental conditions. It is
seen that $I_{max}$ decreases with $E_{J0}$, which in turn is
proportional to the area of the junction. The circles in the inset
show the fit to this universal dependence made using three fitting
variables: the fluctuation free critical current density $J_{c0}$
(the same for all mesas), the effective noise temperature
$T_{eff}$ for the initial mesa and $T_{eff}$ for trimmed mesas
(the same for both since they were measured in the same run). The
horizontal dashed lines in Fig. 2 show $J_{c0}$ (the top line) and
the most probable switching current densities for the three mesas,
obtained from such a fit. Good agreement is seen between fitted
and measured switching currents for all three mesas. Taking into
account that there was no free fitting parameters (three points
were fitted with three variables) and that $T_{eff} =10K$ and
$12K$ are similar and only few degrees above the substrate
temperature $T \simeq 6K$, we conclude that the observed decrease
of the switching current density in smaller mesas is the result of
enhanced thermal fluctuation caused by reduction of the Josephson
coupling energy $E_{J0}$ with junction area.

\begin{figure}
\noindent
\begin{minipage}{0.48\textwidth}
\epsfxsize=.9\hsize \centerline{ \epsfbox{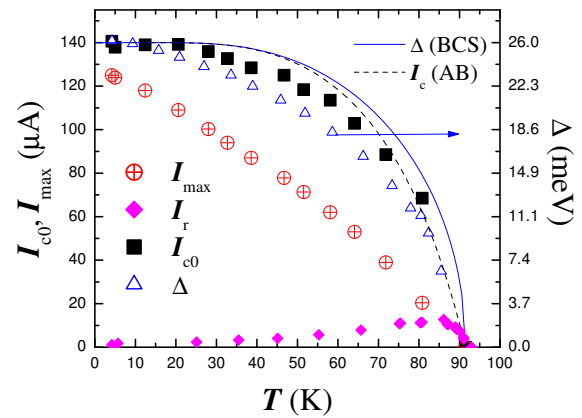} }
\caption{Temperature dependencies of the most probable switching
current $I_{max}$ (circles), the retrapping current $I_r$
(rhombs), the extracted fluctuation-free critical current $I_{c0}$
(squares), and the superconducting gap $\Delta$ (triangles) for
the same IJJ as in Fig. 3. It is seen that $I_{max}$ has unusual
linear dependence in the whole $T-$range. However, $I_{c0}(T)$ is
normal, close to $\Delta(T)$ and consistent with tunnelling nature
of the interlayer transport. For comparison, $T-$dependencies of
the energy gap $\Delta (BCS)$ and the critical current $I_c (AB)$
for conventional SIS junctions are shown by solid and dashed
lines, respectively. }
\end{minipage}
\end{figure}

The thermal escape rate, Eq.(1), strongly depends on the shape of
the wash board potential, $\Delta U (\varphi)$, which in turn
depends on the current-phase relation $J_s(\varphi)$. Therefore,
the probability distribution of the switching current $P(I)$
contains explicit information about the shape of Josephson
potential $E_J(\varphi)$. Since Fig. 2 b) indicates that the
switching current density of small IJJ's is well described by
thermal activation from the wash-board potential, we should be
able to probe the intrinsic Josephson potential by studying
switching current statistics. Unfortunately, previous studies of
the switching current statistics in Bi-2212 mesas revealed that
switching current histograms of mesas containing several stacked
IJJ's may be very unusual. It was reported that histograms of
stacked IJJ's may contain multiple peaks \cite{Mros,Compar} and
appeared to be extremely broad \cite{Mros,Compar,Warburton}, up to
$\sim$ten times broader than expected. Such anomalous behavior was
attributed to the presence of multiple metastable states (fluxon
modes), which appear due to coupling of junctions in the stack
\cite{Modes,Compar}. The existence of metastable states results in
a multiple valued critical current and dramatic enhancement of
thermal fluctuations in stacked Josephson junctions. Note that
such behavior is not specific to Bi-2212 mesas, but was also
observed for low-$T_c$ stacked Josephson junctions \cite{Compar}.

In order to study the intrinsic Josephson potential we have to
avoid metastable states. To solve this problem here we have
studied switching of a {\it single} IJJ. The stable switching of a
single junction can be obtained when there is a spread in critical
currents between IJJ's in the mesa. Such a spread is often
observed in mesas obtained by wet chemical etching, see Fig. 2,
and is most probably caused by variation of the junction area due
to undercut.

Fig. 3 represents the switching current statistics of a single IJJ
in an optimally doped Bi-2212 mesa. This junction had $\sim 20\%$
smaller critical current than the rest of IJJ's in the mesa, which
was sufficient for achieving stable biasing without switching the
rest of IJJ's. Measurements were done in a shielded room
environment using a sample-and-hold setup with the effective noise
temperature $\sim 100 mK$ \cite{Thilo}. Fig. 3 b) shows switching
current histograms obtained from 10240 switching events at
different $T$. The solid lines represent fits to classical thermal
escape, Eqs. (1-3). The fit was made using following parameters:
the experimental sweeping rate $dI/dt=24.3 mA/s$, the specific
capacitance $C=68.5 fF/\mu m^2$, and the junction resistance, $R$,
extracted from the high bias resistivity $\rho_c = 25 \Omega cm$.
Since the probability distribution is only slightly dependent on
$C$ and $R$, those parameters were fixed during the fit to avoid
ambiguity. The only remaining fitting parameter was the effective
"escape" temperature $T_{esc}$, which in the absence of noise
should coincide with $T$. The values of $T_{esc}$ obtained from
the fit are plotted as a function of $T$ in Fig. 3 c). It is seen
that for $T < 72 K$, $T_{esc}$ follows $T$. This, confirms the
validity of the fitting procedure and clearly demonstrates the
sinusoidal current-phase relation and the cosinusoidal dependence
of the intrinsic Josephson potential $E_J(\varphi)$. The
remarkable accuracy with which the sinusoidal current-phase
relationship is satisfied can be seen from the excellent quality
of the fit, which is shown in detail in inset to Fig. 3 b). Such
behavior is characteristic for high quality SIS tunnel junctions.
This is the first unambiguous evidence for the tunnelling nature
of interlayer transport in Bi-2212.

At high temperatures, $T > 75 K$, $T_{esc}$ starts to decrease and
eventually vanishes close to $T_c \simeq 93 K$. The surprising
collapse of thermal fluctuations close to $T_c$ is associated with
the change in the shape of switching histograms, which lose their
characteristic asymmetric form and become symmetric and narrow. It
is also associated with the decrease and collapse at $T \sim 80K$
of the hysteresis in IVC, as seen from comparison of the most
probable switching current, $I_{max}$, and the retrapping current
$I_r$, shown in Fig. 4. We note that the switching current remains
sharply defined and there is no indication for a phase-diffusion
\cite{Muller} up to $\sim 90 K$. We believe that the collapse of
$T_{esc}$ is caused by entering high dissipation regime close to
$T_c$, in which the spread in switching currents is reduced by
enhanced probability of retrapping of the rolling particle in the
wash board potential. We emphasize that such unusual behavior is
not unique for Bi-2212 IJJ's but was also observed in low-$T_c$
superconductor- normal metal- superconductor junctions. Therefore,
this phenomenon is not essential for the present work and will be
discussed elsewhere \cite{Thilo}.

Fig. 4 shows temperature dependence of the most probable switching
current $I_{max}$ (circles), the retrapping current $I_r$
(rhombs), and $I_{c0}$, obtained from fitting switching current
histograms (squares), for the same IJJ as in Fig. 3. It is seen
that $I_{max}$ has an unusual linear dependence in the whole
$T-$range. However, the $T-$ dependence of $I_{c0}$ is quite
normal and close to $T-$dependence of the superconducting gap
$\Delta$ (triangles), which was obtained from the sum-gap knee in
IVC's at higher bias \cite{Krasnov_TH}. For comparison,
$T-$dependencies of the conventional BCS energy gap, $\Delta
(BCS)$ and the Ambeokar-Baratoff value of the critical current
$I_c (AB)$ for conventional SIS junctions are shown in Fig. 4 by
solid and dashed lines, respectively. It is seen that $I_c (AB)
\propto \Delta (BCS)$ at $T < T_c/2$. The experimental $\Delta
(T)$ deviates somewhat from $I_{c0} (T)$ in the intermediate
$T-$range. The deviation is likely a result of self-heating at the
large sum-gap voltage $V=2N\Delta/e$, where $N$ is the number of
IJJ's in the mesa. Such deviation is in agreement with both
numerical simulations \cite{HeatPhC} and in-situ measurement of
self-heating in our mesas \cite{Insitu}. Therefore, the unusual
$T-$dependence of the switching current is solely due to thermal
fluctuations, while $T-$dependence of the extracted fluctuation
free critical current is consistent with the tunnelling nature of
interlayer transport.

In conclusion, having studied the switching current statistics of
a single Bi-2212 intrinsic Josephson junction, we observed that it
can be very well, and without fitting parameters, described by
thermal activation from a tilted wash-board potential with the
sinusoidal current-phase relation. This is direct evidence for the
dc-intrinsic Josephson effect and the first unambiguous
confirmation of tunnelling nature of the interlayer transport in
strongly anisotropic HTSC. We demonstrated that thermal
fluctuations dramatically affect properties of small IJJ's,
resulting in strong suppression of the switching current density
and unusual $T-$dependence in the whole $T-$ range. However,
fluctuation-free $I_{c0}$, extracted from the analysis of
switching current histograms, exhibit a $T-$dependence typical for
SIS tunnel junctions, also confirming tunnelling nature of the
interlayer coupling.

The work was supported by the Swedish Research Council, grant Nr:
621-2001-3236. We are grateful to R.Gross for providing the
sample-and-hold equipment.


\begin{references}

\bibitem{Interlayer} D.G.Clarke, S.P.Strong and P.W.Andersen, {\em Phys.Rev.Lett} {\bf 74}, 4499 (1995);
W.Kim and J.P.Carbotte, {\em Phys.Rev.B} {\bf 63} (2001) 054526

\bibitem{Kleiner} R.Kleiner, et.al, {\em Phys.Rev.Lett} {\bf 68}, 2394
(1992); R.Kleiner and P.M\"{u}ller, {\em Phys.Rev.B} {\bf 49},
1327 (1994)

\bibitem{Fiske} V.M.~Krasnov et al., {\em Phys.Rev.B} {\bf 59}, 8463 (1999)

\bibitem{LatyshPhC} Yu.I.Latyshev, et.al., {\em Physica C} {\bf 362}, 156 (2001)

\bibitem{Ooi} S.Ooi, T.Mochiku, and K.Hirata, {\em Phys.Rev.Lett} {\bf 89}, 247002 (2002)

\bibitem{Wang} H.B.Wang et al., {\em Phys.Rev.Lett.} {\bf 87}, 107002 (2001)

\bibitem{Plasma} M.B.Gaifullin, et al., {\em Phys.Rev.Lett.} {\bf 84}, 2945 (2000)

\bibitem{Compar} V.M.~Krasnov, et al., {\em Phys.Rev.B} {\bf 61}, 766 (2000)

\bibitem{RevMP} A.A.Golubov, M.Yu.Kupriyanov, and E.Il'ichev, {\em Rev.Mod.Phys.} {\bf 76}, 411 (2004)

\bibitem{Martinis} J.M.Martinis, M.H.Devoret, and J.Clarke, {\em Phys.Rev.B} {\bf 35}, 4682 (1987)

\bibitem{Grabert} H.Grabert, P.Olschowski and U.Weiss, {\em Phys.Rev.B} {\bf 36}, 1931
(1987); M.B\"{u}ttiker, E.P.Harris, and R.Landauer,{\em
Phys.Rev.B} {\bf 28}, 1268 (1983)

\bibitem{Krasnov_TH} V.M.~Krasnov et al., {\em Phys.Rev.Lett.} {\bf 84},
5860 (2000); {\em ibid.} {\bf 86}, 2657 (2001)


\bibitem{Mros} N.~Mros, et al., {\em Phys.Rev.B} {\bf 57}, R8135 (1998)

\bibitem{Warburton} P.A.Warburton, et al., {\em J.Appl.Phys.} {\bf 95} 4941 (2004)

\bibitem{Modes} V.M.~Krasnov, et al., {\em Phys.Rev.B} {\bf 56}, 9106 (1997)

\bibitem{Thilo} T.Bauch, et al., unpublished

\bibitem{Muller} R.L.Kautz and J.M.Martinis, {\em Phys.Rev.B} {\bf 42}, 9903 (1990);
A.Franz, et al., {\em Phys.Rev.B} {\bf 69}, 014506 (2004)

\bibitem{HeatPhC} V.M.~Krasnov, {\em Physica C} {\bf 372-376},
103 (2002); V.M.~Krasnov et al., {\em J.Appl.Phys.} {\bf 89}, 5578
(2001); {\em ibid.} {\bf 93}, 1329 (2003);

\bibitem{Insitu} V.M.~Krasnov, M.Sandberg and I.Zogaj, {\em cond-mat} {\bf /0410207}

\end{references}
\end{document}